# Self-consistent model of edge doping in graphene

Thomas Garm Pedersen

*Department of Physics and Nanotechnology, Aalborg University, DK-9220 Aalborg Øst, Denmark*
*Center for Nanostructured Graphene (CNG), DK-9220 Aalborg Øst, Denmark*

Dopants positioned near edges in nanostructured graphene behave differently from bulk dopants. Most notable, the amount of charge transferred to delocalized states (i.e. doping efficiency) depends on position as well as edge chirality. We apply a self-consistent tight-binding model to analyze this problem focusing on substitutional nitrogen and boron doping. Using a Green's function technique, very large structures can be studied and artificial interactions between dopants in periodically repeated simulations cells are avoided. We find pronounced signatures of edges in the local impurity density of states. Importantly, the doping efficiency is found to oscillate with sublattice position, in particular, for dopants near zigzag edges. Finally, to assess the effect of electron-electron interactions, we compute the self-energy corrected Green's function.

## 1. Introduction

Doping is essential for device technology based on graphene [1]. Both electronic and electro-optical devices typically require doping to establish e.g. junctions or plasmonic response [2]. Whereas doping via electrostatic gating can be applied using the atomically thin graphene sheet in a capacitor geometry [3] it is also possible to achieve permanent doping via more traditional means such as substitutional chemical doping [4-6]. This approach is quite similar to traditional doping strategies in usual three-dimensional group IV or III-V semiconductors. However, graphene presents special challenges in that the material is gapless. Hence, dopants don't form localized impurity states positioned in the band gap but, rather, hybridize with a broad range of states [7-14]. This calls for special treatment of the impurity problem and, in particular, the doping efficiency, i.e. the amount of charge given up to the delocalized $\pi$ - orbital system.

In nanostructured graphene devices, the situation is complicated further. Here, a significant dependence on the impurity position is expected. Due to the atomic thickness, electronic properties near a graphene edge are typically quite different from the bulk of the sheet [15-20]. Moreover, the chirality (i.e. armchair vs. zigzag) is expected to affect the impurity states near the edge. Experimentally, the edge is expected to be more chemically reactive than the bulk and so dopants will preferentially substitute carbon at edge sites if introduced after nanostructuring [21,22]. In contrast, if a homogeneously doped graphene sheet is subsequently structured using e.g. lithography, no preference for edge doping is expected. The theoretical understanding of edge doping is still quite limited. The impurity calculation is restricted by the large unit cells typically associated with nanostructured graphene. Hence, systems studied at *ab initio* level are limited to e.g. relatively narrow nanoribbons with little separation between dopants [15-18,20] even though interaction effects are known to be long-ranged [14]. On the other hand, tight-binding models are able to handle very large samples with disorder and low impurity concentration [11,19,23].

In the present work, we wish to eliminate all influences of impurity-impurity interactions as well as artificial size effects. Hence, we should ideally study a semi-infinite graphene sheet with a single dopant located in the vicinity of the edge. For practical reasons, however, we restrict the computations to finite width nanoribbons chosen so wide (approximately 300 graphene units) that their central parts are essentially bulk-like. We use an impurity Green's function method to handle nanoribbons containing isolated dopants. Thus, we find well-defined fingerprints of the site-dependence of doping efficiency for such nanostructures with specific edge chirality. Moreover, local densities of states at impurity sites provide spectrally resolved signatures of edge proximity and chirality. We apply the self-consistent tight-binding method elaborated in Ref. [24]. Here, the dopant is modelled as a substitutional impurity in the carbon $\pi$ - system. Non-orthogonality between neighboring $\pi$ - orbitals is retained in order to break electron-hole symmetry as found in *ab initio* models. Thus, the impurity enters the model as a single-site shift of the atomic on-site energy. Hopping and overlap integrals are taken to be approximately equal for bonds with and without impurity participation. Hence, our model corresponds precisely to the one used in Ref. [24] except for the crucial difference that an edge of specified chirality is introduced. We compare the properties of dopants located in the first atomic rows to those located in the bulk material. To this end, the self-consistent Hubbard correction is applied to explicitly determine impurity occupancy for edge and bulk sites. We then find that doping efficiency oscillates with distance from the edge but, especially for zigzag edges, a reduced efficiency is found for the outermost row. We ignore effects of spin-polarization and geometric relaxation near the edges. Spin-polarization is known to affect pristine zigzag edges but the presence of N-impurities will quench the spin imbalance [20]. Finally, we address the influence of electron-electron interactions via the quasi-particle self-energy.

## 2. Theory and methods

We briefly recapitulate the computational approach for the impurity model referring the reader to our previous exposition in Ref. [24] for details and full derivations. Primarily, the $\pi$ - orbitals are found as eigenstates of a Hamiltonian $H = H_0 + H_1$ with

$$H_0(i,j) = \begin{cases} \varepsilon_p & i=j \\ -t & (i,j) \text{ NN} \\ 0 & \text{otherwise} \end{cases}, \quad H_1(i,j) = \begin{cases} \Delta & i=j=l \\ 0 & \text{otherwise.} \end{cases} \quad (1)$$

Here, $\varepsilon_p$ is the carbon on-site energy and the NN term in $H_0$ describes hopping between nearest neighbor (NN) sites $i$ and $j$ with hopping integral $t = 3$ eV. The impurity Hamiltonian $H_1$ is formulated taking $|l\rangle$ as the impurity site. In the bulk material, all sites are obviously equivalent. In edged graphene, different sites are affected by their proximity to the edge and we label the positions according to the designations in Fig. 1 with the black rectangles indicating unit cells. For armchair and zigzag nanoribbons we choose rather wide structures with 598 and 600 atoms pre unit cell, respectively. Below, we will use the central site $l$ = 149 as an approximant for the bulk case. We have tested even wider ribbons and found only negligible changes in the properties of the central atoms.

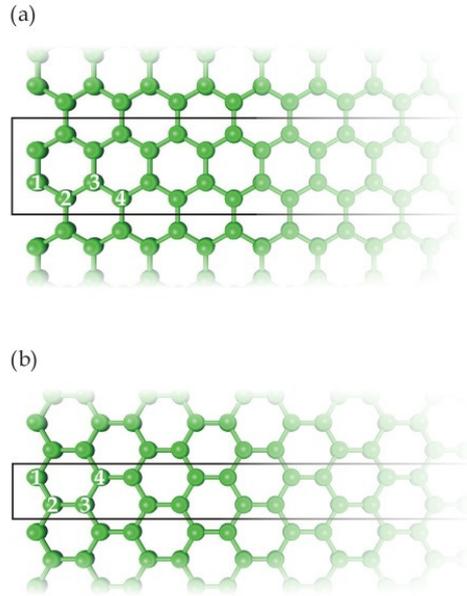

Figure 1. Unit cells of armchair (a) and zigzag (b) edged nanoribbons with numbering of edge sites indicated.

The non-orthogonal eigenproblem for the case at hand is given by

$$(H_0 + H_1) \cdot c_{n,k} = E_{n,k} S \cdot c_{n,k}, \quad (2)$$

where $k$ is the one-dimensional $k$-vector and $S$ is the overlap matrix with unit value at the diagonal and $S(i,j) = s = 0.15$ for nearest neighbors. Based on band structures for the ribbons we form unperturbed Green's functions similarly to the bulk case

$$G_0(z) = \left(z - S^{-1}H_0\right)^{-1}, \quad \tilde{G}_0(z) = \left(zS - H_0\right)^{-1}. \tag{3}$$

In turn, these allow us to construct the diagonal element of the impurity perturbed Green's function

$$G_{ll}(z) = \frac{G_{ll}^0(z)}{1 - \Delta \tilde{G}_{ll}^0(z)}. \tag{4}$$

A crucial observation in the evaluation of these Green's functions is that they follow immediately from the simple one associated with the standard orthogonal problem, i.e.

$$g_{ll}^0(z) = \frac{1}{\Omega_{BZ}} \sum_n \int \frac{\left|c_{n,k}^{(l)}\right|^2}{z - \varepsilon_{n,k}} dk, \tag{5}$$

where $\varepsilon_{n,k}$ and $c_{n,k}$ are eigenvalue and -vector of the standard orthogonal problem $h_0 \cdot c_{n,k} = \varepsilon_{n,k} c_{n,k}$ with $h_0 = -t\sum_{\langle i,j \rangle} |i\rangle\langle j|$ and superscript $l$ indicates projection onto the $l$'th site. In turn [24],

$$G_{ll}^0(z) = \frac{1 + \varepsilon_p \frac{s}{t}}{\left(1 + z\frac{s}{t}\right)^2} g_{ll}^0\left(\frac{z - \varepsilon_p}{1 + z\frac{s}{t}}\right) + \frac{s}{t + zs},$$

$$\tilde{G}_{ll}^0(z) = \frac{1}{1 + z\frac{s}{t}} g_{ll}^0\left(\frac{z - \varepsilon_p}{1 + z\frac{s}{t}}\right). \tag{6}$$

The Green's function of the orthogonal model $g^0$ is computed numerically from the nanoribbon band structure using 200 k-points in the Brillouin zone. In Ref. [25], a semi-analytical method based on image techniques for the evaluation of $g^0$ for semi-infinite graphene sheets with an armchair edge was presented. This method, however, requires computation of off-diagonal elements of $g^0$ and, moreover, cannot readily be generalized to other edge types.

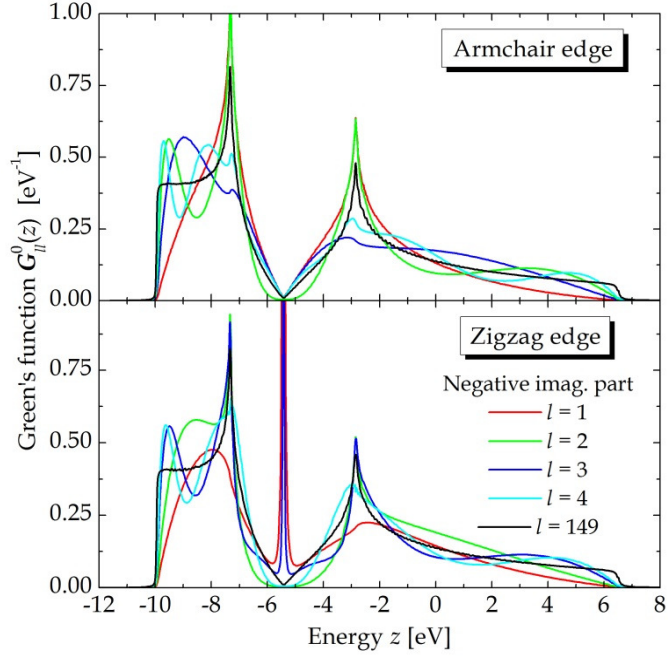

Figure 2. Green's function (negative of imaginary part) versus site for armchair (top panel) and zigzag (bottom panel) edges.

The imaginary part of the Green's function for both edge types is illustrated in Fig. 2 for a range of distances to the edge. For both armchair and zigzag cases, it is clear that the central site ($l$ = 149) behaves very similarly to the case of infinite graphene. The central site Green's functions are practically indistinguishable from the bulk case apart from the slight smoothing due to the small imaginary part (30 meV) added to the energy to avoid numerical instability in the numerical evaluation. In contrast, the behavior near the edge is markedly different. In all cases, the imaginary parts of the Green's functions vanish outside the range $z \in [z_-, z_+]$ with $z_\pm = (\varepsilon_p \pm 3t)/(1 \mp 3s)$. However, for the central site the function jumps to a nonzero value at the limits of the interval while a smooth decrease to zero is found for the near-edge cases. As the distance to the edge increases, the behavior near the limits becomes increasingly steep and oscillations appear. For the zigzag edge, additional deviations are observed near the Dirac point $z = \varepsilon_p$. In this case, the outermost atomic row ($l$ = 1) belongs to a single sublattice, say, the A sublattice. This geometry supports highly localized edge states that lead to the sharp Dirac point feature. A similar behavior is found for other rows belonging to the A sublattice, i.e. for $l$ odd. For even $l$ corresponding to B sublattice sites, no such singularity is observed.

## 3. Impurity properties

The impurity Green's function given by Eq.(4) is readily computed from the unperturbed ones. In turn, the impurity occupancy follows from the integral of the local density of states (DOS) at the impurity site

$$L(\omega) = -\pi^{-1} \operatorname{Im}\left\{\frac{G^0_{11}(\omega)}{1-\Delta \tilde{G}^0_{11}(\omega)}\right\}. \tag{7}$$

Hence, the occupancy including spin degeneracy is

$$n(E_F) = 2\int_{-\infty}^{E_F} L(\omega)d\omega. \tag{8}$$

Here, the upper integration limit is the Fermi level $E_F$. As we introduce only a single impurity in a semi-infinite structure, the Fermi level remains fixed at the Dirac point $E_F = \varepsilon_p$. Generally, the local DOS contains a regular part and a pole contribution [24]. The latter is sharply peaked and, therefore, hard to integrate accurately. However, the normalization $n(\infty) = 2$ can be used to relate the regular and pole contributions and ensure accurate evaluation.

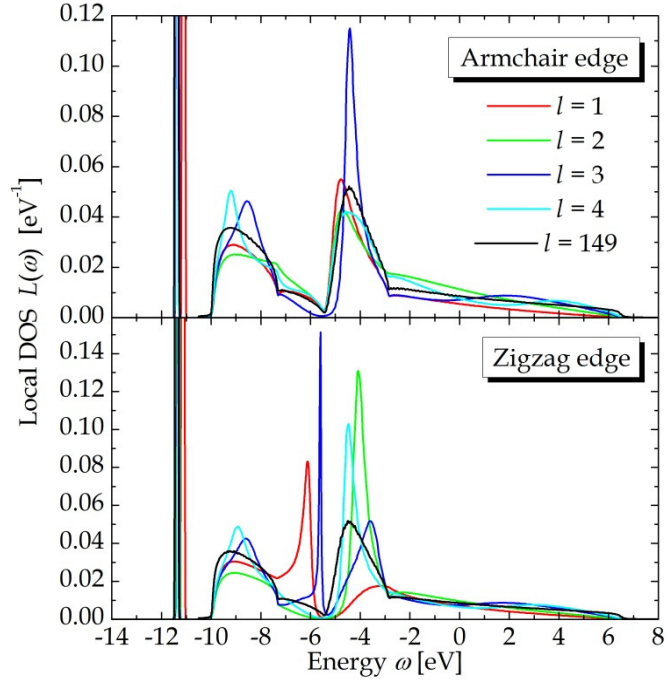

Figure 3. Local density of states for *n*-type impurities (Δ = -5 eV) for various sites and edge types.

In Fig. 3, an example for *n*-type impurities taking $\Delta = -5$ eV is shown. The sharply peaked pole contribution is found near -11.5 eV. The regular part for sites near the edge deviates from the bulk counterpart, in particular, for the zigzag case. It is seen that significant peaks appear in the energy range just below the Dirac point for $l = 1, 3$ and just above for $l = 2, 4$. This again reflects the fact that odd and even sites belong to different sublattices. It is clear that these deviations will influence the impurity occupancy near the edge. For $l$ odd, the increased local DOS below the Fermi level will increase occupancy whereas even $l$ cases will display a reduction. Upon integration to compute the occupancy using Eq. (8) we find the curves in Fig. 4 for a range of impurity potentials and sites. As expected, occupancies tend to oscillate between even and odd $l$ but far more pronounced in the zigzag than armchair case. The curves are roughly symmetrical with respect to inversion around the unit occupancy at $\Delta = 0$ with a slight asymmetry due to the overlap correction breaking electron-hole symmetry. For the outermost site at the zigzag edge, the occupancy varies steeply at $\Delta = 0$. This reflects the very sharp feature in the local DOS that contributes to *n* only for $\Delta < 0$. For $l = 4$, the curves are very close to the bulk case.

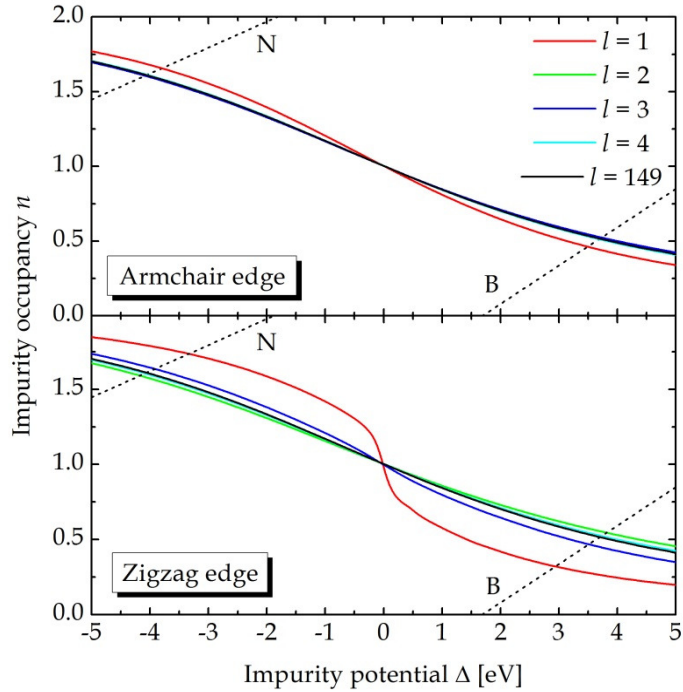

Figure 4. Impurity occupancy for different impurity potentials $\Delta$ and sites. The straight, dashed lines denoted "N" and "B" are derived from the dependence of $\Delta$ on *n* in the self-consistent tight-binding model of these atomic species. Acceptable solutions are given as intersections.

As in the bulk case, we now take advantage of the linearized dependence of $\Delta$ on $n$: $\Delta = \varepsilon_I^0 - \varepsilon_p + U(n - n_0)$, where $\varepsilon_I^0$ is the $2p_z$ energy eigenvalue of a neutral, isolated impurity atom, $U$ is the Hubbard parameter and $n_0$ is the occupancy of the neutral atom. For $U$, we apply the effective value given by half the atomic value [26] computed by the derivative of $\varepsilon_I$ with respect to $n$ at $n=n_0$. Inverting this relation to express $n$ as a function of $\Delta$ leads to the straight lines included in Fig. 4. The intersections between these lines and the numerical occupancy curves correspond to self-consistent solutions to the impurity problem. Results for both dopant species and edge chiralities are shown in Fig. 5. It is clearly observed that site location has an influence on occupancy and, consequently, doping efficiency. The armchair case is rather smooth and only deviates slightly from the bulk value. In contrast, dopants near zigzag edges are characterized by an oscillatory behavior reflecting the sublattice variation. Thus, for nitrogen impurities at a zigzag edge only about 0.26$e$ are delocalized for the outermost position whereas 0.42$e$ is found for second outermost site. Similar figures are observed for hole-doping by B-impurities.

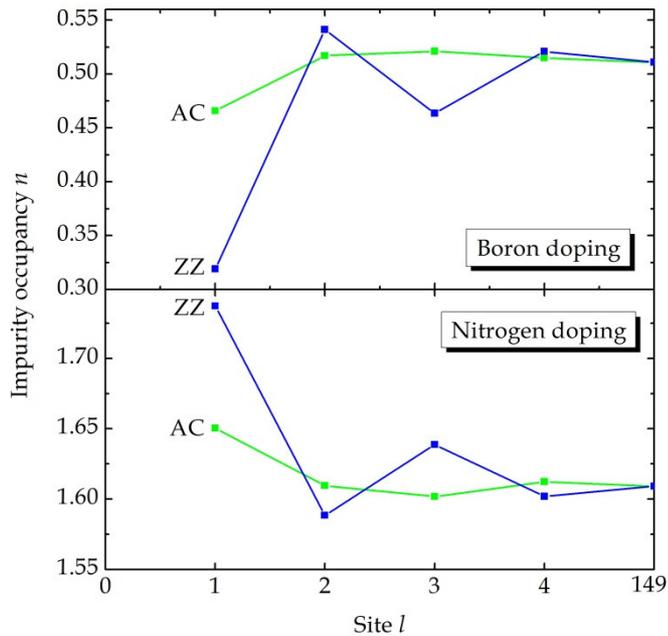

Figure 5. Self-consistent impurity occupancies for N- and B-dopants as a function of positions for armchair (AC) and zigzag (ZZ) edges.

## 4. Electron-electron interactions

The preceeding analysis has focused exclusively on independent electrons. In reality, electron-electron interactions are likely to modify some of the conclusions reached above. In order to quantify these modifications, we now proceed by incorporating interactions in the Green's function. We limit the investigation to the bulk case. The dominant effects of interactions are expected to be described by the complex quasi-particle self-energy $\Sigma$. The real part of $\Sigma$ corresponds to the quasi-particle energy shift, whereas the imaginary part is responsible for the finite lifetime. Generally, the self-energy $\Sigma_n(\vec{k},z)$ depends on both band $n$, energy $z$, and wave vector $\vec{k}$. The relation to the impurity problem is found via the quasi-particle Green's function

$$g^0(z) = \frac{1}{\Omega_{BZ}} \sum_n \int \frac{1}{z - \varepsilon_{n,\vec{k}} - \Sigma_n(\vec{k},z)} d^2k, \tag{9}$$

which replaces Eq.(5). We restrict the analysis to the influence of the finite lifetime on the impurity occupancy determined by the imaginary part of the impurity Green's function. Thus, we focus on the imaginary part of $\Sigma$ and ignore the energy shift produced by the real part. Throughout, we exploit the fact that having a single impurity in semi-infinite graphene amounts to considering intrinsic material. We adopt the zero-temperature ring-diagram approach in the random-phase approximation for the computation of $\mathrm{Im}\,\Sigma$ [27,28]

$$\mathrm{Im}\,\Sigma_n(\vec{k},z) = \frac{1}{4\pi^2} \sum_{m=v,c} \int \left|\langle \varphi_{n,\vec{k}} | \varphi_{m,\vec{k}+\vec{q}} \rangle\right|^2 N(\varepsilon_{m,\vec{k}+\vec{q}},z) \mathrm{Im}\left\{\frac{V_q}{1 + V_q \chi_0(\vec{q}, \varepsilon_{m,\vec{k}+\vec{q}} - z)}\right\} d^2q. \tag{10}$$

Here, $N(\varepsilon,z) = \theta(z-\varepsilon) - \theta(-\varepsilon)$ is the statistical blocking factor and $V_q = e^2/(2\kappa\varepsilon_0 q)$ is the bare Coulomb interaction screened by charges not directly included in the $\pi$- electron model. We take $\kappa \approx 3$ to account for this effect. Also, $\langle \varphi_{n,\vec{k}} | \varphi_{m,\vec{k}+\vec{q}} \rangle$ is the overlap between Bloch states [29] computed using the full tight-binding expression for the wave functions. The susceptibility is

$$\chi_0(\vec{q},\hbar\omega) = \frac{1}{2\pi^2} \sum_{n,m=v,c} \int \left|\langle \varphi_{n,\vec{k}} | \varphi_{m,\vec{k}+\vec{q}} \rangle\right|^2 \frac{f(E_{n,\vec{k}}) - f(E_{m,\vec{k}+\vec{q}})}{E_{n,\vec{k}} - E_{m,\vec{k}+\vec{q}} + \hbar\omega} d^2k. \tag{11}$$

We assume vanishing temperature and evaluate the imaginary part for positive frequencies

$$\mathrm{Im}\,\chi_0(\vec{q},\hbar\omega) = \frac{1}{2\pi}\int \left|\left\langle \varphi_{v,\vec{k}} \middle| \varphi_{c,\vec{k}+\vec{q}} \right\rangle\right|^2 \delta\left(E_{v,\vec{k}} - E_{c,\vec{k}+\vec{q}} + \hbar\omega\right) d^2k \qquad (12)$$

using the improved triangle method [30]. Negative frequencies are handled via $\chi_0(\vec{q},-\hbar\omega) = \chi_0^*(\vec{q},\hbar\omega)$ and the real part is found via the Kramers-Kronig transformation. Due to the symmetry of conduction and valence bands, $\mathrm{Im}\,\Sigma_v(\vec{k},-z) = \mathrm{Im}\,\Sigma_c(\vec{k},z)$. For $z > 0$ it is readily seen that only the conduction band contributes to the sum over bands in Eq.(10). We evaluate Eqs.(9) and (10) by numerical integration over the Brillouin zone and rescale the Green's function to preserve area under the imaginary part.

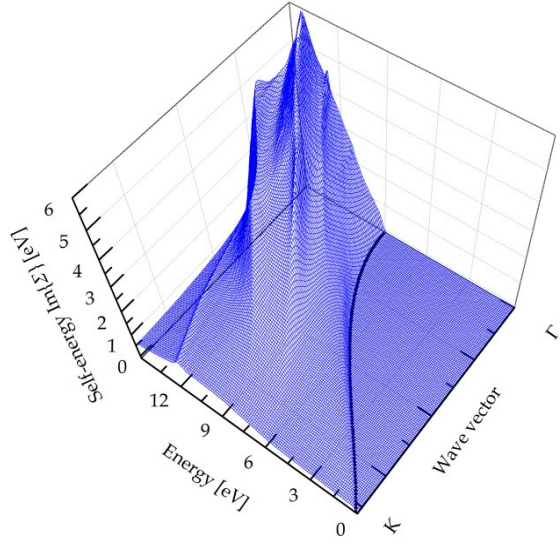

Figure 6. Imaginary part of the conduction band self-energy. The black curve indicates the band edge.

The self-energy evaluated for positive energies is illustrated in Fig. 6. We note that $\mathrm{Im}\,\Sigma_c(\vec{k},z) = 0$ for $z \leq \varepsilon_{c,\vec{k}}$. However, for slightly larger energies, the response increases steeply. Near the K point, the self-energy varies linearly with energy in agreement with the Dirac approximation [27]. Next, we include the self-energy in the computation of the imaginary part of the Green's function using Eq.(6) to incorporate orbital overlap. Since quasi-particle energy shifts are ignored, we only consider the effects of finite lifetime for states in the unperturbed energy range $z \in [z_-, z_+]$ with $z_\pm = (\varepsilon_p \pm 3t)/(1 \mp 3s)$. For energies away from the Dirac point, significant corrections to the Green's function are found as shown in Fig. 7.

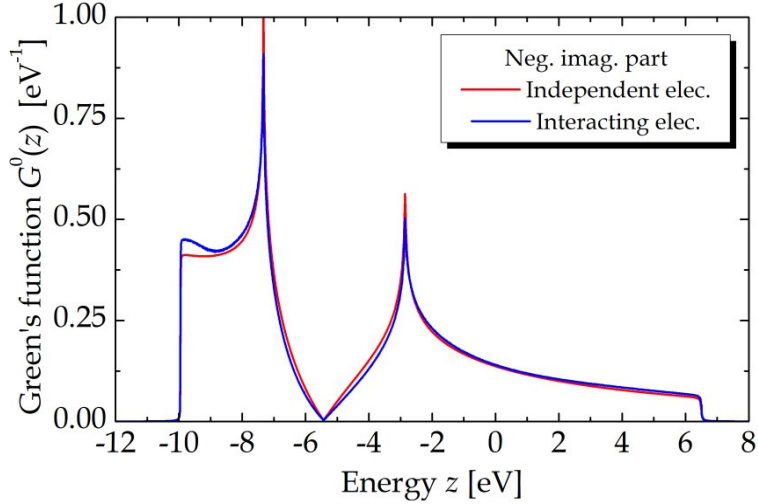

Figure 7. Negative imaginary part of the Green's function with and without electron self-energy effects.

Recalculating the impurity occupancy using the corrected LDOS we find the curve shown in Fig. 8. As expected, little change is observed for impurity levels close to the Dirac point. Notable deviations, however, are found for lower and, in particular, higher energies. For strongly *p*-type impurities, the modified occupancy is higher than the result for independent electrons. Thus, the boron impurity occupancy increases from 0.51 to 0.53 as a result of electron-electron interactions, whereas the nitrogen case remains roughly unchanged.

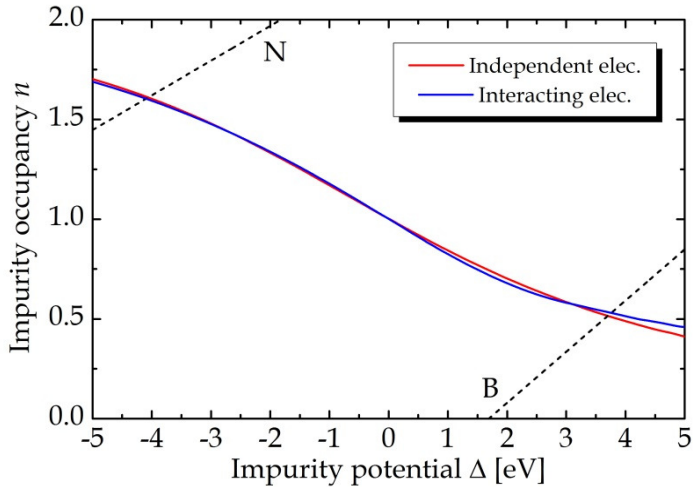

Figure 8. Impurity occupancy with and without electron self-energy effects.

## 5. Summary

In summary, we have studied isolated substitutional dopants in edged graphene using a self-consistent tight-binding framework. Focusing on nitrogen and boron impurities, we have investigated the influence of nearby armchair and zigzag edges. An impurity Green's function approach allows us to compute local densities of states at the impurity site and, thereby, the impurity occupancy. In particular, for dopants close to zigzag edges we find a pronounced oscillatory behavior of this occupancy with distance to the edge. This demonstrates that edges may have a pronounced effect on doping efficiency. In addition, electron-electron interaction effects have been incorporated via the self-energy and we find that *p*-type impurities are most strongly affected.

**Acknowledgments**

This work is financially supported by the Center for Nanostructured Graphene (CNG) and the QUSCOPE center. CNG is sponsored by the Danish National Research Foundation, project DNRF58 and QUSCOPE is sponsored by the Villum foundation.